# Development of high-speed X-ray imaging system for SOI pixel detector

Ryutaro Nishimura

School of High Energy Accelerator Science, SOKENDAI (The Graduate University for Advanced Studies) Tsukuba, Ibaraki, Japan
ryunishi@post.kek.jp

Yasuo Arai, Toshinobu Miyoshi

Institute of Particle and Nuclear Studies, High Energy Accelerator Research Organization (KEK-IPNS) Tsukuba, Ibaraki, Japan
yasuo.arai@kek.jp, tmiyoshi@post.kek.jp

Keiichi Hirano, Shunji Kishimoto, Ryo Hashimoto

Institute of Materials Structure Science, High Energy Accelerator Research Organization (KEK-IMSS) Tsukuba, Ibaraki, Japan
keiichi.hirano@kek.jp, syunji.kishimoto@kek.jp, ryohashi@post.kek.jp

*Abstract*——We are now developing new X-ray imaging system by using Silicon-On-Insulator (SOI) Pixel Detectors. The SOI detector is a monolithic radiation imaging detector based on a 0.2um FD-SOI CMOS process. Special additional process steps are also developed to create fully depleted sensing region of 50~500um thick. SOI detector has up to mega pixels, so development of high speed Data Acquisition (DAQ) system is very important in conducting experiments. We have developed readout board named SEABAS2 (Soi EvAluation BoArd with Sitcp 2) for SOI detector. The SEABAS2 board has 16 channels of 65 MSPS ADCs, 4ch DAC, FPGAs and Gigabit Ethernet I/F. To achieve high throughput of the DAQ, we aggressively adopt parallel processing (data taking and storing) and implement FIFO buffers in software. DAQ throughput in previous DAQ system was 6 Hz (41 Mbps) for INTPIX4 detector which has 423 kpixels of 17 um square. With newly developed system, we could improve this rate to 90 Hz (613 Mbps). To take X-ray images for practical purpose such as 3D CT, user have to control the peripheral devices (moving stage, beam shutter, monitoring devices etc.) while taking images. To ease such data taking, we also implemented automatic control function of peripheral devices. Introduction of SOI detectors, the detail of the DAQ system and experimental results are presented.

## 1 Introduction

X-ray imaging can show the internal structure of an object non-destructively. Especially X-ray imaging with synchrotron radiation is useful for microstructure imaging. For these measurements, the user needs a sensor with high resolution, high detection efficiency and high-speed readout. We have been developing a new-type of monolithic X-ray radiation image sensor by using Silicon-On-Insulator (SOI) technology [1]. The SOI pixel detector is a Si semiconductor detector with small pixel size and can detect charge generated by X-ray. Comparing to in-direct X-ray detector such as using scintillator, the SOI pixel detector can achieve higher spatial resolution (~30um). For practical X-ray imaging, the throughput of the Data Acquisition (DAQ) system and the easy to use are important. In our previous work [2], we presented how to improve the throughput of the DAQ system by refining the DAQ software and we confirmed the throughput increased to 613 Mbps from 41 Mbps. In this work, we present the new framework of whole DAQ system. This new framework improves ease of use by implementation of automatic control of peripheral devices control and realizes extendable application by module structure.

## 2 Specification of SOI pixel detector

SOI pixel detectors are being developed by a SOIPIX collaboration [3] led by High Energy Accelerator Research Organization, KEK. They are based on a 0.2 um CMOS fully depleted (FD-) SOI process of Lapis Semiconductor Co., Ltd [1]. The SOI detector consists of a thick (50–500 um in Floating Zone type wafer) and high-resistivity (more than 2 kΩ-cm in Floating Zone type wafer) Si substrate for the sensing part, and a thin Si layer for CMOS circuits [1]. Comparing to hybrid detector which has thousands of metal bump bonds, the SOI detector has no bump bonds, so the application has low capacitance, low noise, high gain, lower material budget, and can run fast with low power. The structure image is shown in Fig 1.

For the X-ray imaging system, integration-type SOI pixel detectors named INTPIX4 and INTPIX5 are used. The sensitive area is 14.1 x 8.7 mm² (INTPIX4), 16.8 x 10.7 mm² (INTPIX5). These detectors consist of several blocks, and each block has independent channels of analog output for parallel readout. Sensors were mounted on a dedicated sub-board, which was connected to the main DAQ board. In our experiments, INTPIX5 was used when large area and high spatial resolution were required, otherwise INTPIX4 was used. The specification of INTPIX4 and INTPIX5 is summarized in Table 1.

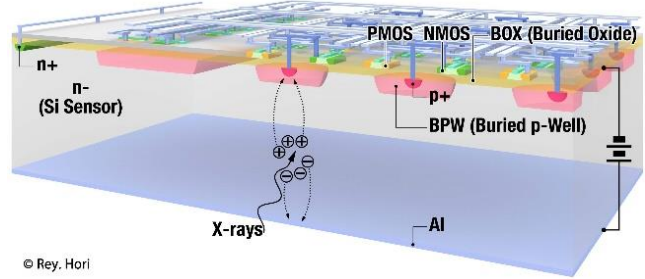

Fig 1. Structure of the SOI detector.

Table 1. The specification of INTPIX4 and INTPIX5

|  | INTPIX4 | INTPIX5 |
|---|---|---|
| Pixel Size (um) | 17 x 17 um² | 12 x 12 um² |
| Number of Pixels | 832 x 512 | 1,408 x 896 |
| Sensitive Area | 14.1 x 8.7 mm² | 16.8 x 10.7 mm² |
| Number of Output channels | 13 (64 x 512pixels / channel) | 11 (128 x 896pixels / channel) |

## 3  SEABAS DAQ system

The SEABAS2 (Soi EvAluation BoArd with Sitcp 2 [4]) is the main DAQ board used to read out SOI sensors. SEABAS 2 is the second generation of the SEABAS board. This board has 16 channels of 65 MSPS ADCs, 4ch DAC, FPGAs and Gigabit Ethernet I/F. The photograph of SEABAS 2 is shown in Fig 2. The signal output by the SOI sensor is an analog signal converted from collected charge. This signal is converted to the count of the A/D conversion Unit (ADU) by the SEABAS2 onboard ADC. After conversion, data are passed to the DAQ PC. The PC and SEABAS2 are connected by Gigabit Ethernet and communicate with TCP / UDP protocol. The transferred data are processed by software that runs on a PC. The schematic of SEABAS DAQ system is shown in Fig 3 and the schematic of data flow is shown in Fig 4.

## 4  Methods of the throughput improvement for SOI sensor DAQ software

### 4.1  Bottleneck in the previous DAQ system

The previous SEABAS DAQ system had the large bottleneck and throughput was limited less than 6 Hz (41 Mbps) for INTPIX4 detector. Here, we predict the cause of the bottleneck. The total throughput depends on the following three points.
1. A/D conversion time. (SOI Sensor - User FPGA).
2. Time of data transfer. (SEABAS2 - DAQ PC).
3. Time of data receiving and data storing. (DAQ PC and inside of DAQ software).

The maximum speed is limited by point 1 because this factor is linked to the hardware (ADC and SOI Sensor) specification. The limitation regarding point 1 is shown in Table 2. In this case, the SOI sensor's performance is the most significant limiting factor. Thus, the required minimum speed of points 2 and 3 is also set to 80 Hz. The total signal length of pixel data is 16 bit/pixel, the width of the digital signal after A/D conversion of the analog signal is 12 bit, and a 4 bit readout channel ID is added to the top of the data stream. Therefore, the total required throughput of 80 Hz readout is 545 Mbps.

Here, we also predict the transfer speed of the SEABAS DAQ system using the Gigabit Ethernet for transfer. When all measurement data are transferred over Ethernet by TCP protocol, the transfer speed that can be used for sensor data is about 95% of the Gigabit Ethernet's maximum speed (1 Gbps). The other 5% is to be used for the structural data of the packet. Furthermore, the prospective transfer speed is approximately 570 Mbps (83.6 Hz)–665 Mbps (97.5 Hz) when the DAQ system can use 60–70% of the theoretical speed. On the other hand, the previous system's transfer speed is 41 Mbps. This means the previous system has some time loss. Most of the time loss is caused by the structure of the DAQ software, in which the processes of data taking and data storage are sequential. Thus, separating the data taking process from other processes will solve this problem.

### 4.2  Implementation of parallel processing and FIFO buffers

In the previous system, the DAQ software works on single-thread processing. In this case, the data taking and data storage processes are sequential. For this reason, the next data taking process has to follow the previous data storage process. Consequently, the net DAQ efficiency is reduced. Therefore, in our works [2], we solved this bottleneck by implementation of parallel processing (data taking and storing) and First In, First Out (FIFO) buffers in the SOI sensor DAQ software. Parallel processing based on multi-thread processing function of WIN32API [5] (Windows) or Posix Thread [6] (Linux). FIFO buffers based on std::list [7]. The schematic of parallel processing and FIFO is shown in Fig 5. In consequence, we could improve throughput from 6 Hz (41 Mbps) to 90 Hz (613 Mbps) for INTPIX4 detector. This performance satisfied the requirement of many experiments.

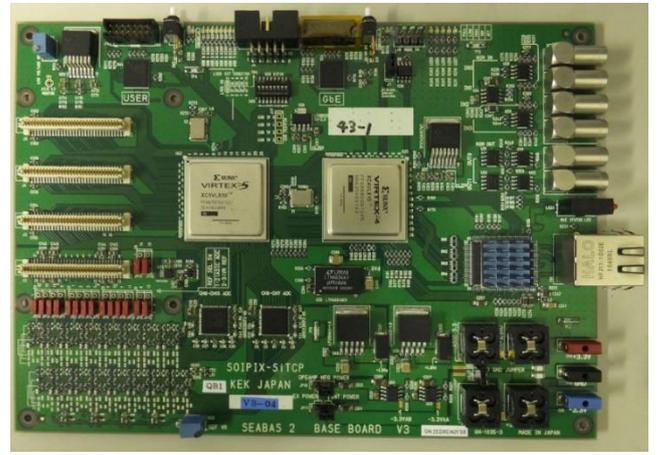

Fig 2. Photograph of SEABAS2.

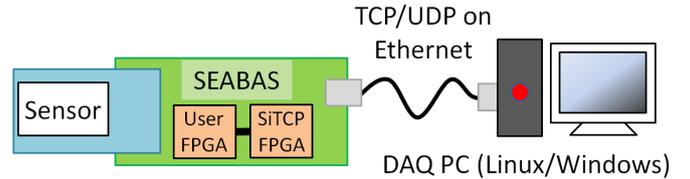

Fig 3. The schematic of the SEABAS DAQ system.

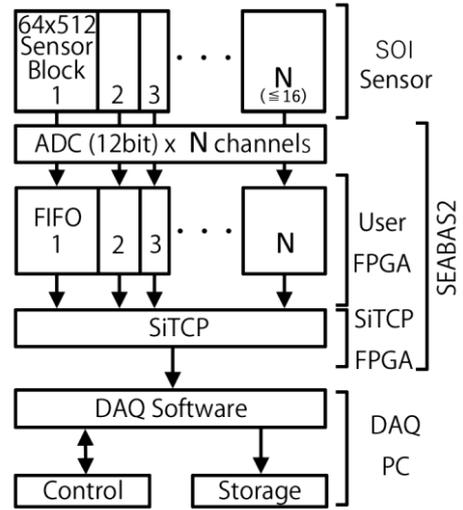

Fig 4. The schematic of SEABAS2 DAQ data flow. (N is the number of SOI sensor's output channels.)

Table 2. Limitation of the SOI pixel detector's frame rate around A/D conversion.

| Component | Value |
|---|---|
| SEABAS 2 A/D Converter | 65 MSPS / channel (Over 1.5 kHz (655.36 µs / frame) of INTPIX4 frame) |
| Sensor's performance (INTPIX4) | 80 Hz (12.5 ms / frame)* |
| * Scan time (switching period) is 320 ns/pixel, and integration time (exposure period) is 2000 µs. ||

## 5 Proposed DAQ framework

### 5.1 Motivation of DAQ framework construction

In our previous work [2], DAQ software had only basic functions at that moment. It's not enough to take X-ray images for practical purpose such as 3D Computed Tomography (CT). To use this DAQ for such purpose, DAQ software must have functions for automatic control of peripheral devices control. Therefore, in this work, we constructed new DAQ framework to extend functions for practical purpose X-ray imaging.

### 5.2 Structure of DAQ framework

Our new DAQ framework is consist of modules which provide simple functions and text-based command connection path. SOI sensor control software, developed in our previous work, also works as one of modules. Basically each modules runs independently. Thus, user can customize DAQ system for their purpose by rearranging modules. In addition, each module doesn't have to have many functions, has to have only a few functions for command connection path. This is effective to reduce source code and to make easier to maintain. This framework can support Windows (Vista or later) and Linux (Cent OS 6 or later), and can be used on mixed environment, for example, some of DAQ PCs using Windows and others using Linux. The schematic of the framework is shown in Fig 6.

## 6 Experimental Results

Proposed DAQ system is tested in KEK Photon Factory [8] BL-14B beam line. In this tests, the setup requires complex DAQ control. Detail of test and results are shown below.

### 6.1 A still absorption image taken by 9.5 keV mono X-rays and reconstruction of 3D CT data.

New X-ray imaging system based on proposed DAQ framework was tested with 3D CT absorption imaging setup. A small dried sardine was used as the sample of 3D CT. In this test, INTPIX4 (N type Floating Zone Si used in sensor layer) and INTPIX5 (N type Czochralski Si used in sensor layer) were used. The measurement conditions are following; INTPIX4's integration time of 2 ms / shot and a total of 1 s (integrate 500 shots) per image, INTPIX5's integration time of 4 ms / shot and a total of 2 s (integrate 500 shots) per image, the bias voltage of the sensor layer 200 V, the scan time 320 ns / pixel, sensor temperature is 20-30 °C (room temperature). X-ray irradiation is done from the backside of the sensor. In this test, X-rays beam energy is tuned to 9.5 keV monochromatic light. The schematic of the setup is shown in Fig 7 and the procedure of this test is listed below.
(a) Rotate the sample around the vertical axis and take still X-ray images every 1 degree, from 0 to 180 degrees.
(b) Take X-ray profile of PF Beam.
(c) Make projection data from (a) and (b).
(d) Reconstruct 3D CT data from (c) by filtered back-projection method.

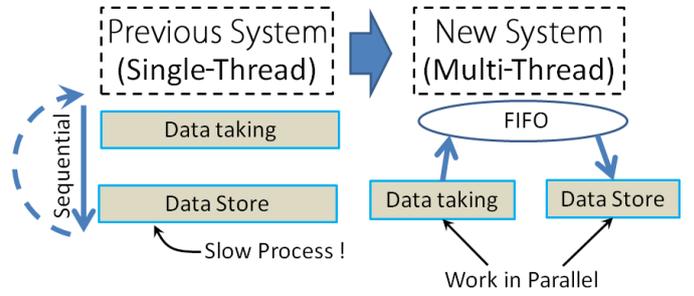

Fig 5. The schematic of parallel processing and FIFO.

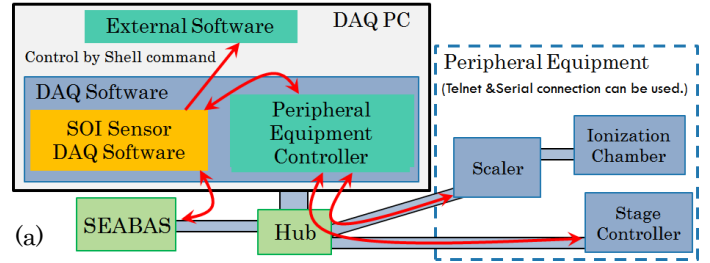

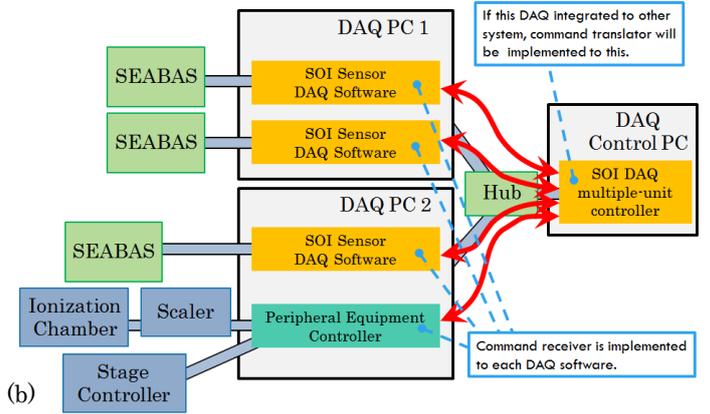

Fig 6. The schematic of DAQ framework. - (a) For single sensor setup.
(b) For multiple sensors setup.

Fig 8 and 9 show samples of the image taken in this test and Table 3 shows the required time of data taking in this test. This imaging system worked very well and took all data without an error. Present dead time is limited by the maximum integration time of the sensor which is determined by sensor leakage current. The leakage current can be reduced dramatically by cooling the sensor. We are prepared cooling system for the next beam test. Then the integration time will be limited only by X-ray intensity and maximum charge stored in a pixel if the cooling system works.

## 7 Conclusion

In this paper, we proposed and built a DAQ system for X-ray imaging system using SOI detector. Our new DAQ framework is consist of functional modules and SOI sensor DAQ module. This module structure makes easier to customize and maintain the system. In addition, this framework support several operating system and can be used even on mixed environment. We applied new X-ray imaging system for the experiment at KEK Photon Factory BL-14B. In this test, the new DAQ took 181 still X-ray images and 1 X-ray beam profile for 3D CT reconstruction automatically and we could reconstruct data with a small number of artifacts. We confirmed performance of our new X-ray imaging system.

## 8 Acknowledgements

This work was supported by MEXT KAKENHI Grant-in-Aid for Scientific Research on Innovative Areas 25109001, and it was performed with the approval of the Program Advisory Committee of the Photon Factory (2013G054, 2014G021 and 2015G649). This study was performed as part of the SOIPIX group [3] research activity.

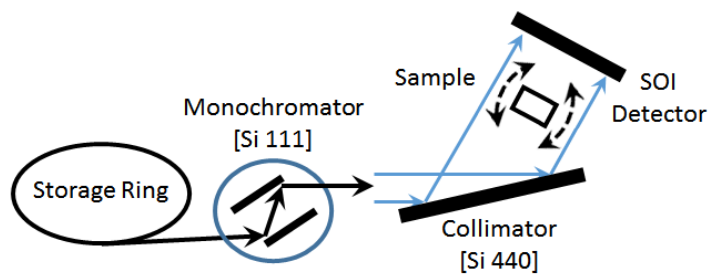

Fig 7. The schematic of the setup for absorption imaging 3D CT experiment.

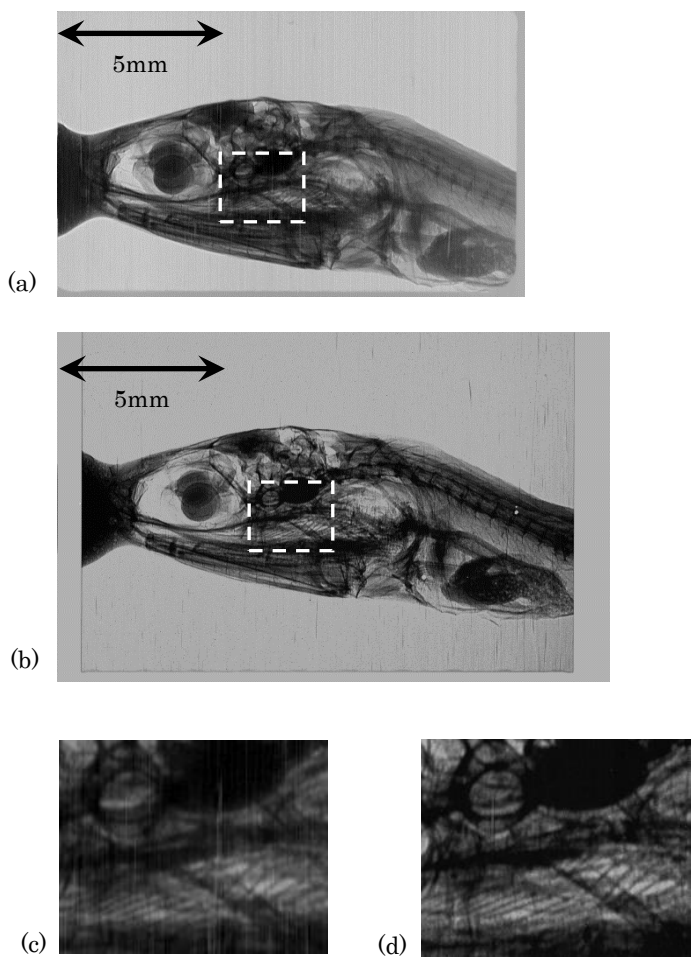

Fig 8. Still X-ray images of same sample taken by SOI sensors - (a) Taken by INTPIX4. (b) Taken by INTPIX5. (c) Expand image of region framed by white dashed line in (a). (d) Expand image of region framed by white dashed line in (b).

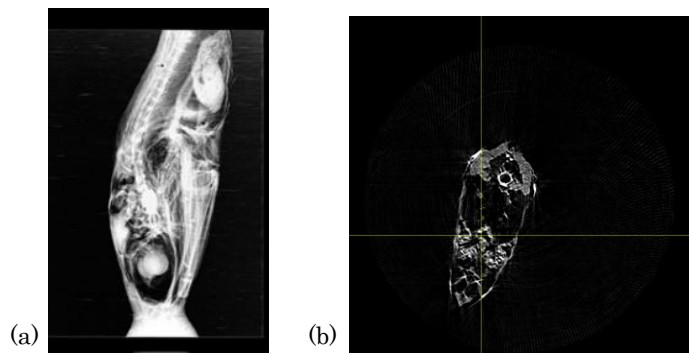

Fig 9. Reconstructed 3D CT Data - (a) One of X-ray projection images made from X-ray images taken by INTPIX5. (b) One of slices reconstructed from X-ray projection images taken by INTPIX5.

Table 3. Required time of data taking for 3D CT experiment.

|  | INTPIX4 | INTPIX5 |
|---|---|---|
| Total data taking time (s) | 4,079 | 7,874 |
| Total integration time (s) | 181 (1 s / image x 181) | 362 (2 s / image x 181) |
| Total time of peripheral equipment control (s) | 2,179 (12 s / image) | 1,357 (7-8 s / image) |
| Dead time (s) | 1,719 | 6,155 |
| ** Dead time : Remaining time excluded integration time and peripheral equipment control time from total data taking time. | | |